\documentclass[a4paper,12pt,onecolumn]{article}

\usepackage[dvips]{graphicx}

\begin{document}
\begin{flushleft}
\Large
\textbf{Network properties, species abundance and evolution in a model
of evolutionary ecology. \\}
\vspace{1cm}

\renewcommand{\thefootnote}{\fnsymbol{footnote}}
\large
\textbf{Paul Anderson and Henrik Jeldtoft Jensen\footnote[1]
{Author for correspondence (h.jensen@imperial.ac.uk; 
URL: http://www.ma.imperial.ac.uk/$\sim$hjjens)} \\}
\normalsize
\textit{Department of Mathematics, Imperial College, 180 Queen's Gate, London SW7 2AZ, U.K. \\}\renewcommand{\thefootnote}{\arabic{footnote}} 

\end{flushleft}
\linespread{1}

\vspace{1cm}
%
\begin{center} ABSRTACT \end{center}
\noindent
We study the evolution of the network properties of a populated network 
embedded in a genotype space characterised by either a low or a high number of 
potential links, with particular emphasis on the connectivity and clustering. 
Evolution produces two distinct types of network.
When a specific genotype is only able to influence a few other
genotypes, the ecology consists of separate non-interacting clusters in genotype space. 
When different types may influence a large number of other sites, the network becomes 
one large interconnected cluster. The distribution of interaction strengths --- but not the
number of connections --- changes significantly with time. 
We find that the species abundance is only realistic for a high level of species 
connectivity. This suggests that real ecosystems 
form one interconnected whole in which selection leads to
stronger interactions between the different types.   
Analogies with niche and neutral theory are also considered.

\vspace{.5cm}
\noindent \textbf{Short title:} Network properties.
\section{INTRODUCTION}
An important characteristic of an ecosystem 
is the total set of interactions between the various individuals.
Organisms may influence each other in many ways 
and it is difficult to monitor and quantify all possible interactions except the most 
direct, such as simple trophic relations. The 
development of the set of interactions over evolutionary time scales is
even more difficult to measure because of random
mutations and the resulting adaptations. Gaining an understanding from observations 
is also problematic since {\it laws} may very well only be recognisable at the 
level of {\it averages}, see \cite{lore:part,yedi:macr}. 
Here, we approach these issues within the framework of a 
simple model of ecosystem assembly and evolution
\cite{chri:tang,hall:time,diCo:tang}. 

We compare early and late time connectivity and cluster properties of ecosystems evolving 
in two differently connected spaces: genotypes influence 
either a small or a large number of others. 
Clearly, the actual number of interactions experienced 
by a site depends on which of all the possible mutations and adaptations 
have occurred, i.e. the network is dependent on its history. It turns
out that the interaction strengths change significantly with time,
whilst the degree (number of active interactions) distribution remains close to what would be expected
if genotypes were occupied at random.  The species abundance curve
takes a log-normal form only for spaces where the genotypes are linked
to many others. Our model is neutral in some aspects but also draws on
concepts from niche theory.

\section{METHODS}
We briefly describe the structure and dynamics of the Tangled Nature
model. Details can be found in \cite{chri:tang,hall:time}.
An individual is represented by a vector ${\bf S}^\alpha=
(S_1^\alpha,S_2^\alpha,...,S_L^\alpha)$ in the genotype space
$\cal{S}$, where the ``genes'' $S^\alpha_i$ may take the values $\pm 1$, 
i.e. ${\bf S}^\alpha$ denotes a corner of the 
$L$-dimensional hypercube. In the present paper we take $L=20$. 
We think of the genotype space $\cal{S}$ as containing all possible ways
of combining the genes
into genome sequences. Many sequences may not correspond to viable
organisms. The viability of a genotype is determined by the
evolutionary dynamics. All possible sequences are made available for evolution to select from.
The number of occupied sites is referred to as the diversity.

For simplicity, an individual is removed from the system with a
constant probability $p_{kill}$ per time step.   
A time step consists of {\em one} annihilation attempt followed by
{\em one} reproduction attempt. One generation 
consists of $N(t)/p_{kill}$ time steps, which is the average time 
taken to kill all currently living individuals. All references to time
will be in units of generational time.

The ability of an individual to reproduce is controlled by a weight
function $H({\bf S}^\alpha,t)$:  
\begin{equation}
H({\bf S}^\alpha,t)={1\over c N(t)} \left( \sum_{{\bf S}\in{\cal S}} 
J({\bf S}^\alpha,{\bf S})   n({\bf S},t) \right)
- \mu N(t),
\label{Hamilton2} 
\end{equation}
where $c$ is a control parameter, $N(t)$ is  the total number of individuals at time $t$, 
the sum is over the $2^L$ locations in ${\cal S}$ and $n({\bf S},t)$ is 
the number of individuals (or occupancy) at position ${\bf S}$. 
Two positions ${\bf S}^a$ and ${\bf S}^b$ in genome space are coupled with the fixed random 
strength $J^{ab}=J({\bf S}^a,{\bf S}^b)$ which can be either positive, negative or zero. 
This link is non-zero with probability $\theta$. There is no
self-interaction, so $J^{aa}=0$. The present paper compares
the three cases $\theta=\frac{1}{1000}$, $\theta=\frac{1}{200}$ and $\theta=\frac{1}{4}$. 
The non-zero values of $J^{ab}\neq J^{ba}$ are determined by a
deterministic but rapidly varying function of the two 
positions ${\bf S}^a$ and ${\bf S}^b$ \cite{hall:time}. 

The conditions of the physical environment are simplistically
described by the term $\mu N(t)$ in equation (\ref{Hamilton2}), 
where $\mu$ determines the average sustainable total population size, i.e. 
the carrying capacity of 
the environment. An increase in $\mu$ corresponds to harsher physical
conditions. Notice that genotypes only adapt to each other and the
physical environment represented by $\mu$. 
We use asexual reproduction consisting of one individual
being replaced by two copies.
Successful reproduction occurs with a probability per unit time given by
\begin{equation}
p_{off}({\bf S}^\alpha,t)={ \exp[H({\bf S}^\alpha,t)]\over
1+\exp[H({\bf S}^\alpha,t)]}\in[0,1].
\label{p_off}
\end{equation}
We allow for mutations in the following way: with probability $p_{mut}$ per gene we perform a 
change of sign $S_i^\alpha \rightarrow - S_i^\alpha$, during reproduction. 

Initially, we place $N(0)=500$ individuals at randomly chosen positions. 
Their initial location in genotype space does not affect the
dynamics. A two-phase switching dynamic is seen consisting of long
periods of relatively stable configurations (quasi-Evolutionary Stable Strategies
or q-ESSs) interrupted by brief spells of reorganisation of occupancy
which are terminated when a new q-ESS is found, as discussed in \cite{chri:tang}.

\section{RESULTS}

\renewcommand{\thesubsection}{\alph{subsection}}
\subsection{\textit{Parameters}}
As mentioned previously, the only parameter that is changed in this
paper is the connectivity, $\theta$. The values used throughout
are $L=20$, $c=0.005$, $\mu=0.01$, $p_{kill}=0.2$ and
$p_{mut}=0.015$. This selection provides several transitions between different
q-ESSs for each run, see \cite{hall:time,diCo:tang}. 
We consider three $\theta$ values: $0.001$, $0.005$ and $0.25$, which we will refer to
as very low, low and high $\theta$ respectively.
These correspond to below, near and above the percolation
threshold. That is, the point where there is a \textit{non-zero} probability
that all living sites are connected
in one dominant cluster \cite{albe:stat}. A realistic species abundance curve was only obtained
above the threshold.
We will be contrasting results at $t=500$ (primal time), $t=5\ 000$
(early time) and $t=500\ 000$ (late time). 
Early time is well outside the system's initial transient search for a
quasi-stable configuration in genotype space.  
The low and high $\theta$ ensembles consist of 500 realisations. Each run 
uses a different random number seed but, for any given
run, only $\theta$ is changed between the two ensembles.
 
Each figure in the paper, apart from figure \ref{fig4}, shows two sets of data:
one labelled \textit{simulation} (which are the results generated by the
dynamics of the model) and the other \textit{random}. In the random case, rather
than evolving the network, for any specified time we read in the diversity 
and number of individuals alive in the simulated run. 
The individuals are then thrown on to the network of $2^{20}$
genotypes at random with the constraint that the diversity is the same
as the simulation. Thus, random data is \textit{not} dependent on the history
of the network, but has the same global properties (diversity and
population size) as the simulations. This provides a very useful null model.
Comparisons with this procedure will reveal
whether the network is really evolving, or the results are just
by-products of increasing diversity. Simulated data is always shown as a 
dotted line and random as a continuous line.

\subsection{\textit{Connectivity}}
\begin{figure}[p]
\begin{center}
\includegraphics[width=1.0\linewidth]{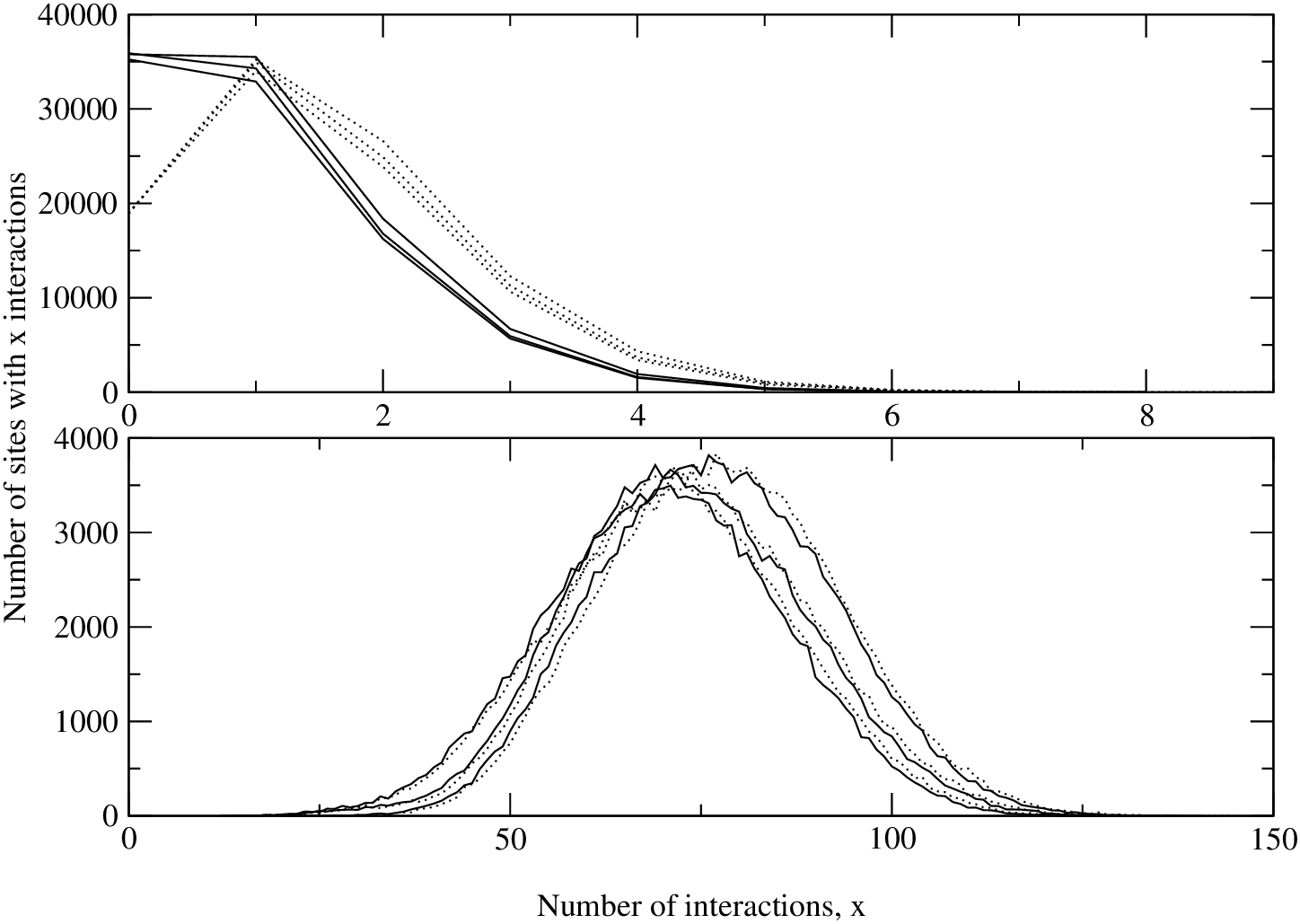}
\caption{Top: Degree histogram for $\theta=0.005$. Bottom:
$\theta=0.25$. Solid lines, random; dotted lines, simulation. 
From the left, the pairs of curves are for t=500, 5000
and 500000. At later times, the number of active links increases for both the 
simulation and random data.}
\label{fig1}
\end{center}
\end{figure}

We study the temporal evolution of the network connectivity in the
space of occupied positions for different $\theta$ values.  
Note that the hard-wired configuration of
couplings $J({\bf S^a},{\bf S^b})$ between all $2^{20}$ positions in genotype
space is determined at $t=0$ and remains constant. The network of {\em
occupied} sites will nevertheless change with time. 
The degree distributions in figure \ref{fig1} show the number of genotypes
having $x$ \textit{active} interactions.

The leftmost pair of curves represents
primal time, the next, early time and the rightmost late time. 
Considering only the simulation data for now, a clear shift to a
greater number of active links is seen in the high
$\theta$ case, whilst a slight change occurs for low $\theta$. 
The difference between early and late time is bigger than that between
early and primal time. The degree of a site is equal to the number of 
direct interactions it has with all other occupied sites.
This explains why any particular site in the low $\theta$
runs only has at most nine and usually only one or two direct
interactions. The data is integrated over
each ensemble. 
How much of this shift is due to a genuine change in network
connectivity? For high $\theta$, the null model data shows that there is very little
difference between evolving the network and throwing down individuals
randomly. Low $\theta$ appears to show a change. However, any site that does not
interact with any others will die very quickly in a
simulation. If for any instant in time genotype positions are
chosen by chance, such a low connectivity will give a disproportionate
number of isolated genomes that would be forbidden by the dynamics. 
There is no fair way to simulate this
effect, but it can be seen that the differences between the time
curves in the random and simulated runs is similar and thus 
the network connectivity does not evolve for either value of $\theta$.
   
\subsection{\textit{Interaction strength}}

\begin{figure}[p]
\begin{center}
\includegraphics[width=1.0\linewidth]{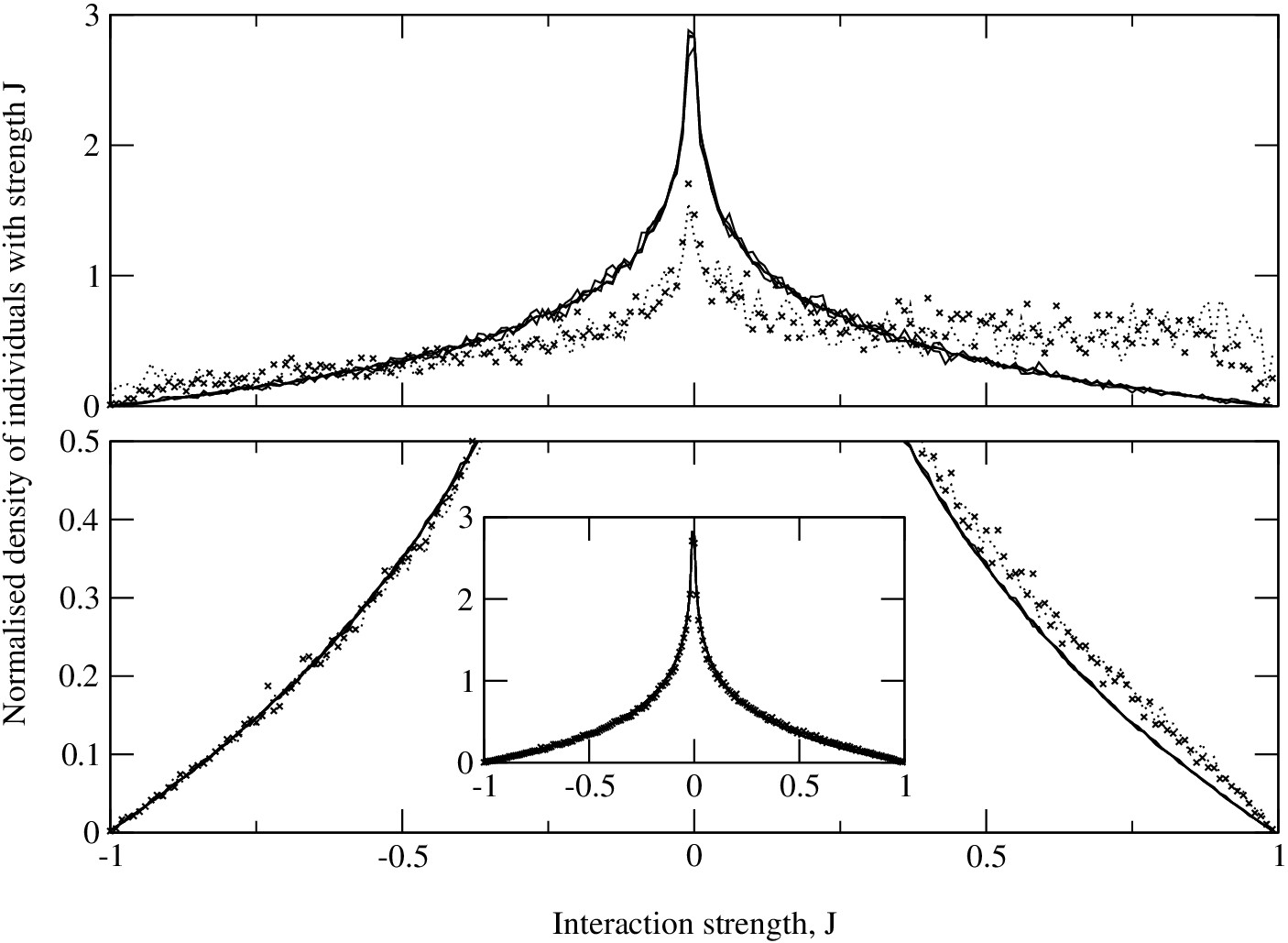}
\caption{Top: Distribution of interaction strengths between
individuals for
$\theta=0.005$. Bottom: $\theta=0.25$. Inset: Entire distribution.
Solid lines, random; crosses, simulation at t=500; dotted lines,
simulation at t=500000.
All plots are normalised so that their area is one.
For high $\theta$, a significant increase in
positive interactions is seen. For low $\theta$, a change is seen but
for trivial reasons.}
\label{fig2}
\end{center}
\end{figure}

For both $\theta$ values, the diversity gradually increases with time. 
What is causing this? It turns out that 
the strength of the interaction between sites is crucial to the ability
of the network to support larger numbers of individuals. Figure
\ref{fig2} shows the distribution of interaction
strengths between each \textit{living site} and all other
living sites at a given time. Interaction strengths are assigned at
random and are not necessarily symmetric \cite{hall:time}. For example, 
$J({\bf S}^{18},{\bf S}^{59})=0.3$, but $J({\bf S}^{59},{\bf S}^{18})=-0.2$. 
For all times and
both values of $\theta$, the distribution for the random data is a
sharply pointed, symmetrical curve peaking at $J=0$. This makes sense
because there is no bias in the ratio of positive to negative
links when the links are assigned to the ``bare'' network at $t=0$.

For reasons of clarity, the simulation results are only shown for
primal and late time. Clearly, a significant change takes place for
high $\theta$ between $t=0$ and primal time. Some weight is taken in the
fall of the peak at $J=0$ and the drop in negative J values, and
redistributed into positive strengths i.e. the curve shifts
right. This comprises a significant shift in the probability density.
The move from primal to late time
is smaller --- but still noticeable --- since the large number of
reorganisations of genotype space in the early generations drops to
occasional punctuation of q-ESSs later in the run. (Typically, there
are only one or two transitions from early time onwards.) Despite this,
the curve continues to drift to the right. 
On first inspection, the low $\theta$ runs seem to have changed
dramatically from the initial configuration. However, nothing
particularly interesting is happening here: it is simply an effect of
the structure of clusters in the low $\theta$ space and is explained
below.

\subsection{\textit{Clustering}}
The indirect connectivity or clustering (how sites are linked to each other through
other sites) is another useful network measure. For 
high $\theta$ we find that at any given time, all occupied 
sites in genotype space belong to one and the same cluster. Thus the cluster
size for high $\theta$ simply follows the diversity.
In contrast, we observe the formation of distinct clusters for low
$\theta$. The rest of this section will deal solely with low $\theta$. 

The overall structure of the clusters
does not seem to change much with time. This is the first indication
that the clustering is not an evolving property of the network. 
As would be expected, one-clusters are transient. They are born on
a new site as mutants from a parent but are isolated from other
sites and so are extremely unlikely to reproduce. (Since $\mu=0.01$
and the average total population is about 2700, $p_{off} \approx
10^{-12} \ll p_{kill}=0.2$ so an isolated site is much more likely to
be killed than multiply when chosen.) These sites are simply flashing
in and out of existence.

Simulations indicate that the building blocks of larger groups are 
two-clusters. These tend to be two very old sites that have mutually
positive links. Large clusters are formed mainly from very old
two-clusters joined together by a mutant. 
The continual background of mutants flitting in
and out of the network plugs these building blocks together. However, the
entire cluster is rarely long lived, whereas the two-clusters
are formed early in any particular run, quickly
build up their population and are very persistent since their
occupation is high. 

\begin{figure}[p]
\begin{center}
\includegraphics[width=1.0\linewidth]{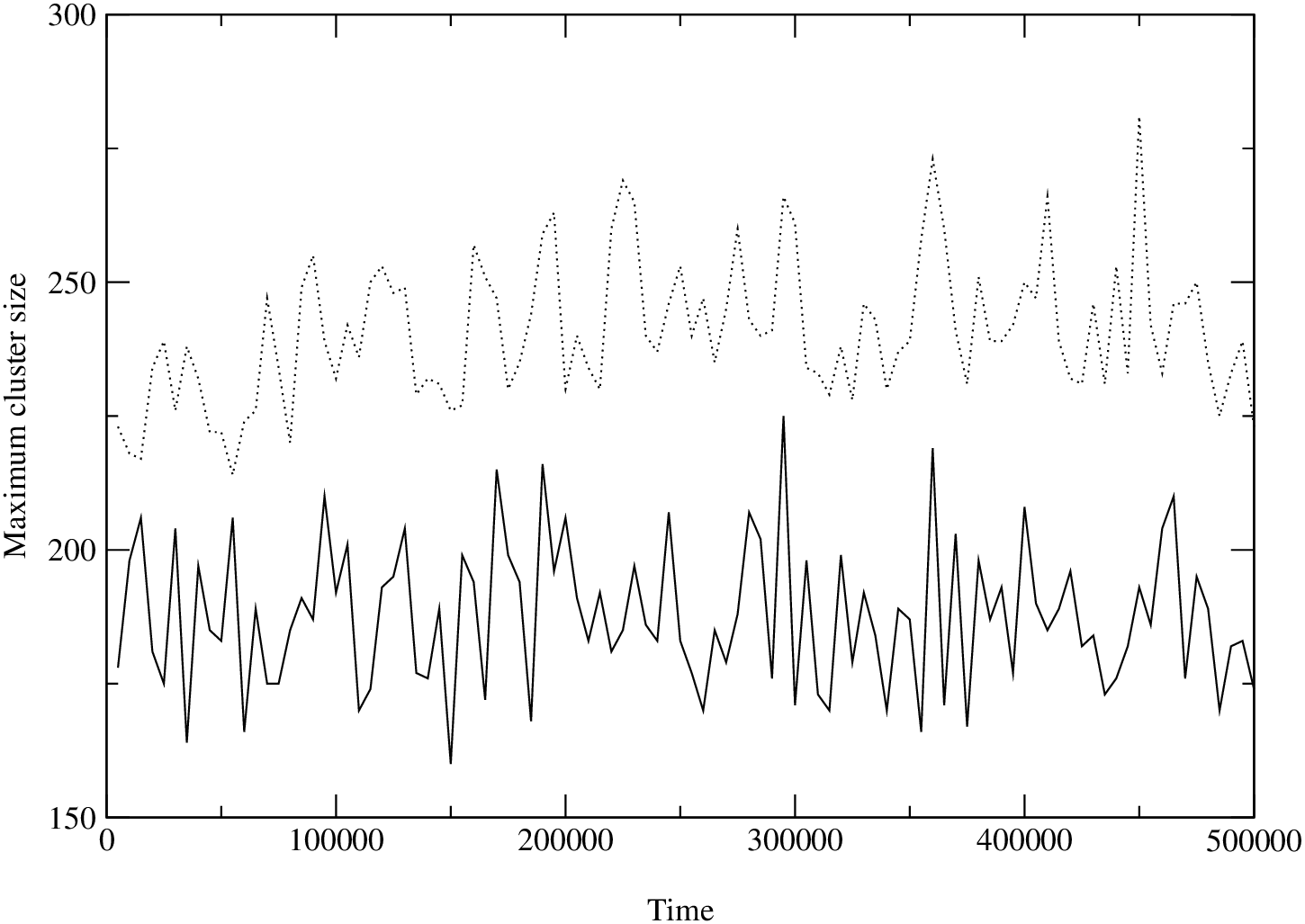}
\caption{Maximum cluster size across all realisations for
$\theta=0.005$. Solid line, random; dotted line, simulation. 
Clusters produced by the simulation are larger than
those produced in a history-independent network}
\label{fig3}
\end{center}
\end{figure}

\begin{figure}[p]
\begin{center}
\includegraphics[width=1.0\linewidth]{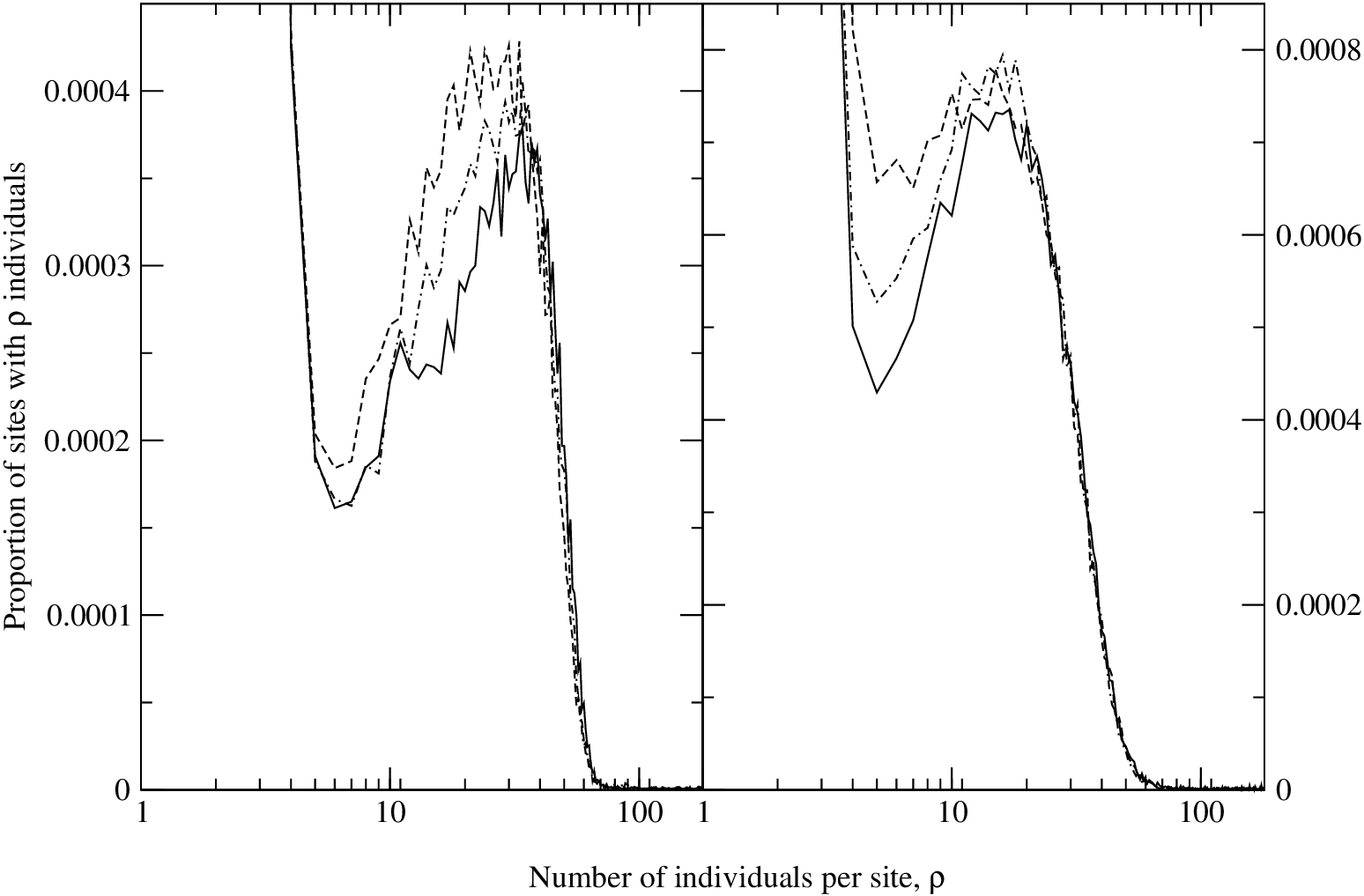}
\caption{Species abundance functions for the simulations only. Dashed line, t=500;
dashed-dotted line, t=5000; solid line, t=500000.
Low $\theta$ on the left, high $\theta$ on the right. The ecologically
realistic log-normal form is only seen for high $\theta$.}
\label{fig4}
\end{center}
\end{figure}

Clusters do indeed generally increase in size with time. 
There are, however, large
fluctuations in the record size which gives an indication of how
unstable these large clusters are, see figure \ref{fig3}. 
The largest recorded cluster in any run at any time contained 281
sites. It is revealing to compare the results from the null model, where
the maximum cluster size is much smaller but just as variable. When
individuals are thrown down at random, two-clusters
are no longer the building blocks of the large clusters and any long
string of connected sites is determined purely by chance and hence the
biggest cluster will always be smaller than that produced by the dynamics.
The temporary nature of the large clusters is further borne out by 
time and ensemble averaging the cluster sizes. The number
of clusters of a particular size $S$ is stored at intervals of 5000
generations from early to late time for each of the 500 runs and the
time and ensemble average is then calculated. As
expected, there are many one-clusters and fewer large clusters.

The distribution follows the functional
form $n_{s} \sim s^{-5/2} e^{-(p-p_{c})^{2} s}$ anticipated for the
cluster size distribution on a random graph of $D$ ($\approx 195$) nodes.
(See equation (36) in \cite{albe:stat}.) The percolation threshold
$\frac{1}{D}$ is very close to the considered connectivity $\theta=0.005$.
A comparison with the random data shows that this scale-independent
distribution is not due to the dynamics
of the system, the only difference being the appearance of larger
record clusters
in the simulation, as shown in figure \ref{fig3}. This is
perhaps the most compelling piece of evidence that the low $\theta$
regime does not show any emergent structure.
We also ran simulations for very low $\theta$ ($0.001$) and found that
the cluster sizes were exponentially distributed as would be expected
below the percolation threshold.

Figure \ref{fig2} (low $\theta$) can now be easily explained. Unconnected sites die
extremely quickly in the first few generations leaving behind
two-clusters and other sites with positive interactions. 
The slight increase in sites with $J>0$
for late time is caused by well established positive-positive
two-clusters. So what looked like an interesting result initially
proves to be due to the fairly constant microscopic structure of the network.

\subsection{\textit{Species abundance}}
The Species Abundance Distribution or SAD is important in 
characterising ecosystems. It is the proportion of species that 
contain $\rho$ individuals. 
We define a species as one site in genotype space. Ideally, we would
like to use a coarse-grained definition more likely to reflect real
ecologies, where species are defined as groups of points in genotype
space echoing the genotypic cluster species definition introduced
by Mallet \cite{mall:spec}. Since the maximum number of genotypes in our
model is only around $10^{6}$ anyway, the single site species approach
is more appropriate. We have been able to extend the initial results
obtained in \cite{hall:time} and can consider the evolution of the
SAD for high and low $\theta$ integrated across
all 500 runs, as seen in figure \ref{fig4}. 
The larger ensembles allow enough statistics for illuminating conclusions to be drawn. 
Note that the null model is absent since when individuals are sprinkled
randomly across the living sites, there is no tendency for
accumulation on any particular site, so the individuals follow a multinomial distribution.

The key result of this paper is that only high $\theta$ leads to a SAD
similar to those observed in nature. Low $\theta$ is skewed by its
heavily populated two-clusters. The plots for high $\theta$ show 
the log-normal form observed in many real ecosystems
and in other ecological models, see \cite{mcKa:mean,hubb:unif}. 
They appear to become more log-normal as time increases with the dip
between four and eight individuals falling, even though the diversity
is rising. Hence, the SAD is evolving. 
From this, it seems that the high $\theta$ case structures
itself more like a real ecosystem than low $\theta$, whose SAD
develops a sharper peak as the two-clusters become densely
occupied. The single cluster of highly interdependent genomes
produces a reasonable SAD that cannot be formed by patches of
isolated clusters. 

Thus the abstract parameter $\theta$, which cannot be
measured in a simple way in real systems, is directly linked to the easily observed SAD.
We recall that low values of $\theta$ correspond to a world in which different
species, or types, are able to influence only
a small number of other species. High values of $\theta$ correspond to
the situation where different types may have an impact on the vitality of a large number
of other species.

The initial descent in both curves from the global peak at 
$\rho=1$ is due to the large number of
sites with only one occupant. In nature, sampling difficulties would
mean that these sites would not be detected so this first aspect is
not seen in observed SADs. (It is particularly
marked for our model since we use each site as one species and do not
coarse-grain.) But the second peak does correspond well to
results from the field, though it should be pointed out that the
proportion of all sites with more than two individuals is only about
$30\%$ in each case. However, this is sufficient to detect the
evolution of the SAD.

We note that a recent study of a simplified version of the Tangled Nature
model by Rikvold and Zia \cite{Rikvold1,Rikvold2} found no temporal evolution of the statistics
of the model. The reason for this may be that they
use a relatively short genome length $L=13$ together with a very substantial
simulation time of order $10^7$ generations. We have observed previously
that the time to reach a stationary state explodes with genome length 
\cite{chri:tang}.

\subsection{\textit{Neutral and niche theory}}
There has recently been much interest in the neutral theory of
biodiversity \cite{hubb:unif}, 
\cite{bell:neut}. Despite making assumptions that are anathema to
traditional niche models (all individuals are the same and adaptations
to specific environmental niches are essentially unimportant) it has
been successful in making predictions about real world ecology ---
although its effectiveness in modelling the species abundance has recently been
called into question \cite{mcgi:ates}.

At $t=0$, the Tangled Nature model is neutral as all individuals
are the same. However, the dynamics immediately breaks this
neutrality as configurations are spontaneously generated. Once
individuals become differentiated, interactions matter and the
evolution is better described by niche-theory --- although $\mu$ and
$p_{kill}$ remain neutral. Unlike the models in \cite{hubb:unif}, we
have no spatial aspect: we deal with only one large metacommunity as opposed to many local
communities aggregating to form the metacommunity. 
 
The only measure in our model considered by the neutral theory is the
species abundance, which we find takes an approximate log-normal
form. The shape predicted by neutrality, the zero-sum multinomial (ZSM)
distribution, is quite similar to the log-normal except that the ZSM
has a long right tail and is much harder to calculate
\cite{mcgi:ates}. To detect whether our distributions are closer to
the ZSM or the log-normal would be computationally prohibitive as the
genome length $L$ would need to be much larger.
Perhaps in the future, models like Tangled Nature could be used to investigate the
relative importance of niche and neutral effects.

\section*{Discussion}
Our most important results are that 
temporal evolution of the network properties of an ecosystem and a
realistic form for the species abundance are only seen if 
the genotype space is well connected. This is interpreted here as meaning that an occupied genotype
is likely to interact with many other (potentially occupied) genotypes.
No evolution at the level of ecosystems can occur
in a world where most possible genotypes are \textit{inert},
i.e. whether they are present or not will have very little influence
on other organisms. It is easy to overlook the importance of the entire network of 
interactions when dealing with small communities of organisms on a
macroscopic scale, but easier to visualise with colonies of billions
of bacteria.

We suggest that this observation can be used to gain insight
into the potential underlying connectivity between biota. Imagine two microbial
evolution experiments. In one case, the microbial ecosystem
evolves towards an interwoven or entangled ecology. In the other, little
evolution is observed in the structure of the ecological properties of
the microbial community. One might, according to the result from our model,
anticipate that the first system consists of microbes from a part
of genotype space in which types influence each other, whereas the
second system consists of genotypes from a region of space
consisting of mainly \textit{inert} organisms.

From our results, it is tempting to speculate that the observed degree
of diversity, complexity and adaptation of living matter may be directly related
to a high level of interdependence between organisms.
Thus Darwin's entangled bank may be a useful image to keep in mind 
when studying the evolution of large
collections of individuals.\\

We are grateful to Gunnar Pruessner for extensive help with the
development of the parallelised code that enabled us to probe the model
at much deeper timescales and for broader ensembles than had 
been possible before. We wish to thank Andy Thomas for his fantastic technical
support.  Without his help and dedication, this project would not have
been possible.  We are indebted to  Dan Moore, Brendan Maguire and
Phil Mayers for their continuous support.
We thank Albert Diaz-Guilera for inspiring discussions
and Ed Johnson for reading the manuscript.
Paul Anderson thanks EPSRC for research funding.

\bibliographystyle{apalike}
\bibliography{complex}

\end{document}